\documentclass[12pt]{article}
\usepackage{graphicx}
\input{epsf}
\begin{document}

\title{Steps in creation of educational and research web-portal of nuclear knowledge BelNET}
\author{
S.Charapitsa{$^1$}, I.Dubovskaya{$^2$}, I.Kimlenko{$^2$}, \\
A.Kovalenko{$^1$}, A.Lobko{$^1$}, A.Mazanik{$^1$}, N.Polyak{$^1$}, \\
T.Savitskaya{$^2$}, S.Sytova{$^1$}\thanks{E-mail:sytova@inp.bsu.by}, A.Timoschenko{$^2$}\\
{\small \it {$^1$}Research Institute for Nuclear Problems of Belarusian State University}\\
{\small \it {$^2$}Belarusian State University, Minsk, Belarus}}
\date{}
\maketitle

\begin{abstract}
Belarusian State University is currently developing the
educational and research web portal of nuclear knowledge BelNET
(Belarusian Nuclear Education and Training Portal). In the future,
this specialized electronic portal could grow into a national
portal of nuclear knowledge. The concept, structure and taxonomy
of BelNET portal are developed. The requirements and conditions
for its functioning are analyzed. The information model and
architecture of the portal, as well as algorithms and methods of
software are realized. At present, BelNET software implemented all
the basic functions of this portal, including the ability to
remotely (via the Internet) open content editing, sorting,
filtering, etc. Filling the BelNET by knowledge is at the
beginning.
\end{abstract}

\section {Introduction}
The International Atomic Energy Agency (IAEA)
\cite{IAEA1}--\cite{IAEA4} pays close attention to the problems of
nuclear knowledge management. Nowadays, numerous national and
international portals of nuclear knowledge were created in Europe,
Asia, Africa and America under the patronage of the IAEA. It is
planned to develop an international network of information
resources on nuclear knowledge. This means that a unified
information space in the field of nuclear knowledge is being
formed in the world. Every developed country with its own nuclear
industry has to create and maintain a national portal of nuclear
knowledge, integrated into a global system of nuclear knowledge
management. In the light of creation of Belarusian nuclear
industry and construction of Belarusian nuclear power plant,
development of electronic portal of nuclear knowledge is an
insistent need for Belarus.

The development of computer technology, new requirements for the
volume, complexity and speed of information transfer, as well as
the rapid growth of mobile applications with specific requirements
on the amount and form of presentation of information demand new
effective algorithmic, architectural and software solutions. The
portal of nuclear knowledge should be a complex programming system
based on such modern technologies. Also, in the light of rapid
growth of popularity of free software in the world, it would be
good if the portal was developed on the Belarusian free software.
So, creating such a portal is not just the development of a simple
website, like such as millions in the Internet. The portal must
meet the requirements of safety, reliability, efficiency and
performance and reflect the national features of nuclear knowledge
content.

The educational and research web-portal of nuclear knowledge
BelNET ({\bf Bel}arusian {\bf N}uclear {\bf E}ducation and {\bf
T}raining) has the following objectives: acceleration of search
and access to necessary data and information, creation of new
knowledge, promotion of participation in research, education and
training programs in nuclear industry, management of information
resources, knowledge and competencies of nuclear industry in
Belarus.

The basis of BelNET software is ``Electronic document management
system eLab''\cite{MMA} developed on free software by the
Laboratory of analytical research of the Research Institute for
Nuclear Problems of Belarusian State University. eLab is
implemented in the educational process of leading Belarusian
universities: Belarusian State University, Belarusian State
Technological University, Belarusian National Technical
University. It is introduced in the Chemical-toxicological
laboratory of the Minsk Drug Treatment Clinic. eLab has been a
basis of management of specimens, measurements and passports of
fuels and lubricants of Belarusian Army since 2012 and Belarusian
branch of Russian company GazPromNeft since 2013. Software eLab is
protected by four Certificates of the National Intellectual
Property Center of the Republic of Belarus.

eLab software is an electronic system of client-server
architecture, designed on the basis of free software: Debian GNU /
Linux, Web-server Apache, the Firebird database server using the
application server PHP. It works under Windows and Linux through
widely used browsers reliably without interruption. It is also
completely secure from unauthorized access. eLab has a fast
response to user requests, providing visibility and accessibility
of information through a single interface for a wide range of
integrated applications for users with different rights of access.
It proved to be a system easily upgradable to conditions of the
project.

\section {BelNET principles of operation}
\begin{figure}[htb]\centering{
{\includegraphics[scale=0.35]{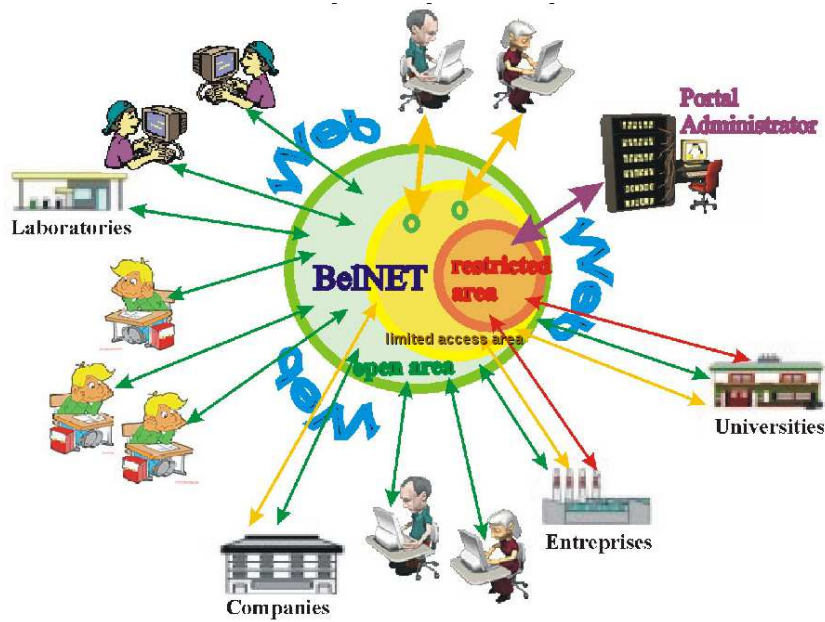}}
 \caption{BelNET principles of operation}\label{fig1}}
\end{figure}

In Fig.\ref{fig1}, the main principles of BelNET operation are
depicted in a simple form. Users with different rights of access
from high school and university students to university professors
and specialists visit on-line BelNET portal from their computers
and laptops. Depending on the rights of access they can read
documents available in open access, limited access or restricted
access areas. With a user name and a password with appropriate
rights, one can enter new documents on-line, or edit the existing
ones etc.

\begin{figure}[htb]\centering{
{\includegraphics[scale=0.9]{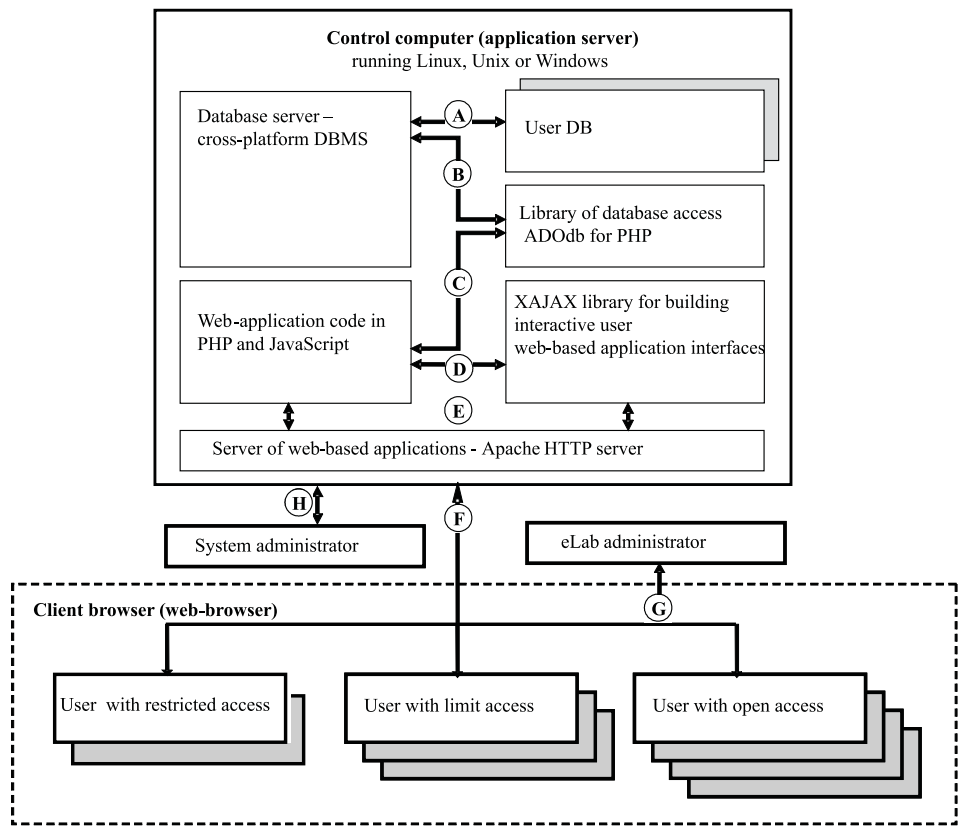}}
 \caption{eLab system architecture}\label{fig2}}
\end{figure}

\begin{figure}[htb]\centering{
{\includegraphics[scale=1.]{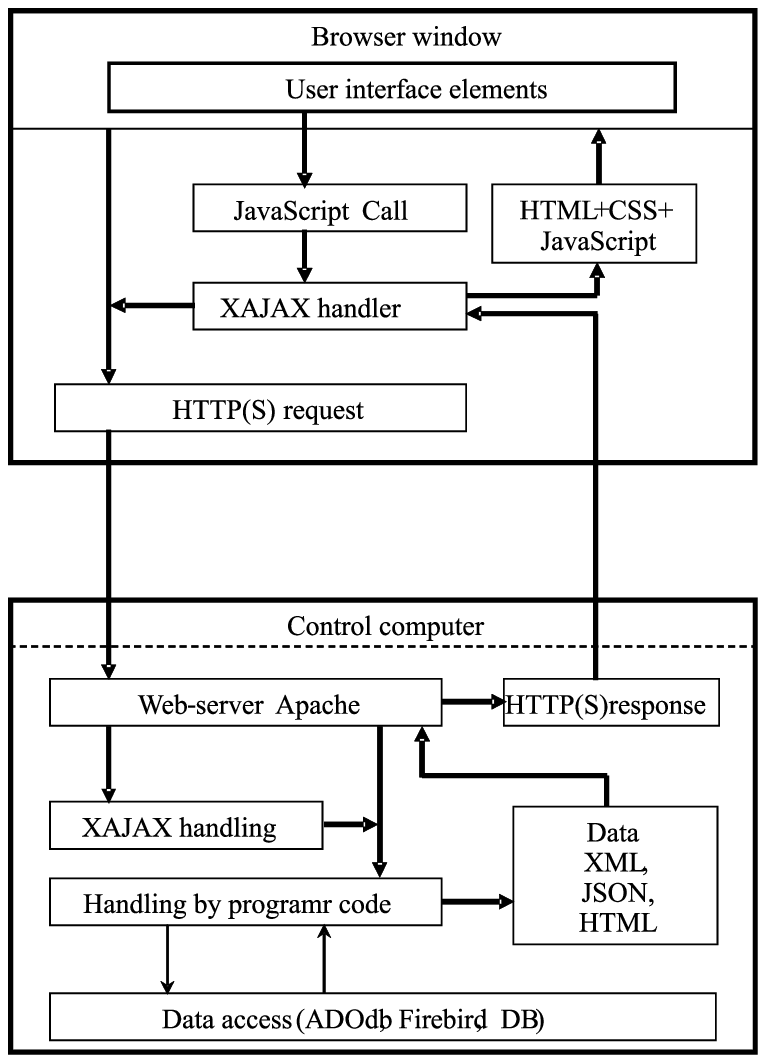}}
 \caption{eLab data streams}\label{fig3}}
\end{figure}

BelNET components presented in Fig.\ref{fig2} as eLab system
architecture are as follows:

1) Control computer (application server) running Linux, Unix or Windows;

2) Database server -- cross-platform database management system (DBMS) Firebird;

3) User databases (DB);

4) Server of web-applications -- Apache HTTP Server;

5) Code of web-application in PHP and JavaScript;

6) Library of access to databases ADOdb for PHP;

7) XAJAX library for building interactive user interfaces and web-applications;

8) Explorer (web-browser of client) for the following user categories: user with restricted
access, user with limited access, user with open access;

9) eLab administrator;

10) System administrator.

Data streams in eLab are the following:

A. Data streams between the database server Firebird and user DB;

B. Data streams at ADOdb library and the DBMS Firebird;

C. Interaction of web-based applications eLab with user DB through ADOdb library and the Firebird;

D. Implementation of XAJAX library in web-application eLab;

E. Formation and processing of HTTP(S) requests by server of web-applications Apache;

F. Data transfer from the server to the client and back through the web-server of applications;

G. Interaction of eLab administrator with eLab system;

H. Interaction of system administrator with the application server
and eLab.

The interaction of these data streams (from Fig.\ref{fig1}) are
presented in other form in Fig.\ref{fig3}.

BelNET and data content are placed on the host computer of the
network (the application server) by the system administrator. The
system administrator has a full and direct access to the
application  server, including  BelNET with its databases. The
system administrator is responsible for operation, safety, and
protection of server applications and data.

BelNET users, including the administrators of web-application
(portal administartors) are clients of the system. They interact
with the system and data over the Internet or the internal
(corporate) network through a browser that is installed and used
on the user's workstation. Personal desktop computers, laptops,
tablets, or smartphones can be used as a workstation. Data streams
between clients and web-application in both directions are carried
out via the web-server Apache which provides validation, filtering
and redirection of HTTP(S) requests. An interactive user interface
is generated on the application server and displayed in a browser
window on a workstation via the server (in PHP) and client (in
JavaScript) of the interlayer XAJAX in accordance with the HTTP(S)
user requests. The interface includes custom elements: links,
buttons, lists, tables and other DOM-elements. Dynamic calls to
the server with technology AJAX are supported allowing the
modification of contents of the browser without reloading the
entire page (the window contents).

\section {Pilot version of BelNET}
\begin{figure}[htb]\centering{
{\includegraphics[scale=0.27]{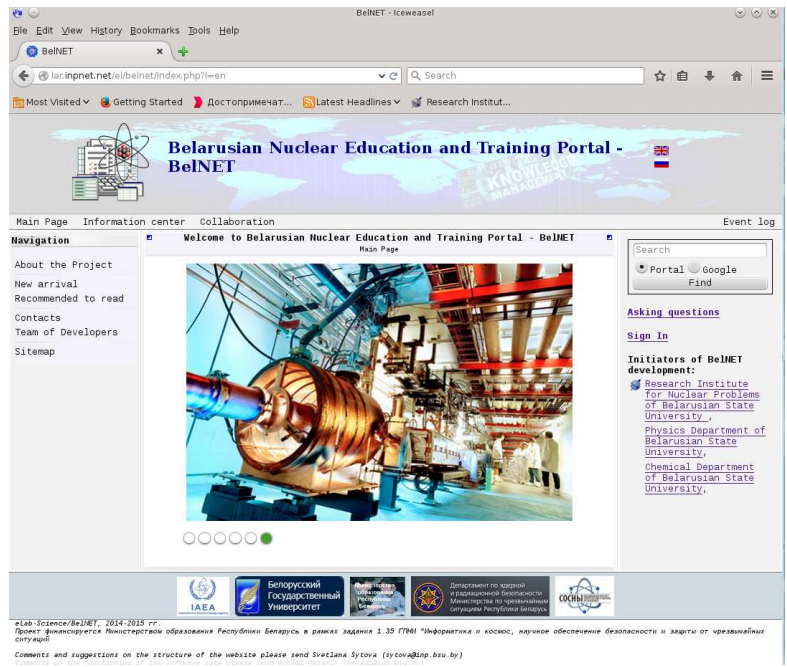}}
 \caption{Start page of BelNET}\label{fig4}}
\end{figure}

The pilot version of BelNET is depicted in part in \cite{lanl2014}--\cite{Congress2015}
and available here:
http://lar.inpnet.net/el/belnet/ (see screenshot of its start page
in Fig.\ref{fig4}).

\begin{figure}[htb]\centering{
{\includegraphics[scale=0.27]{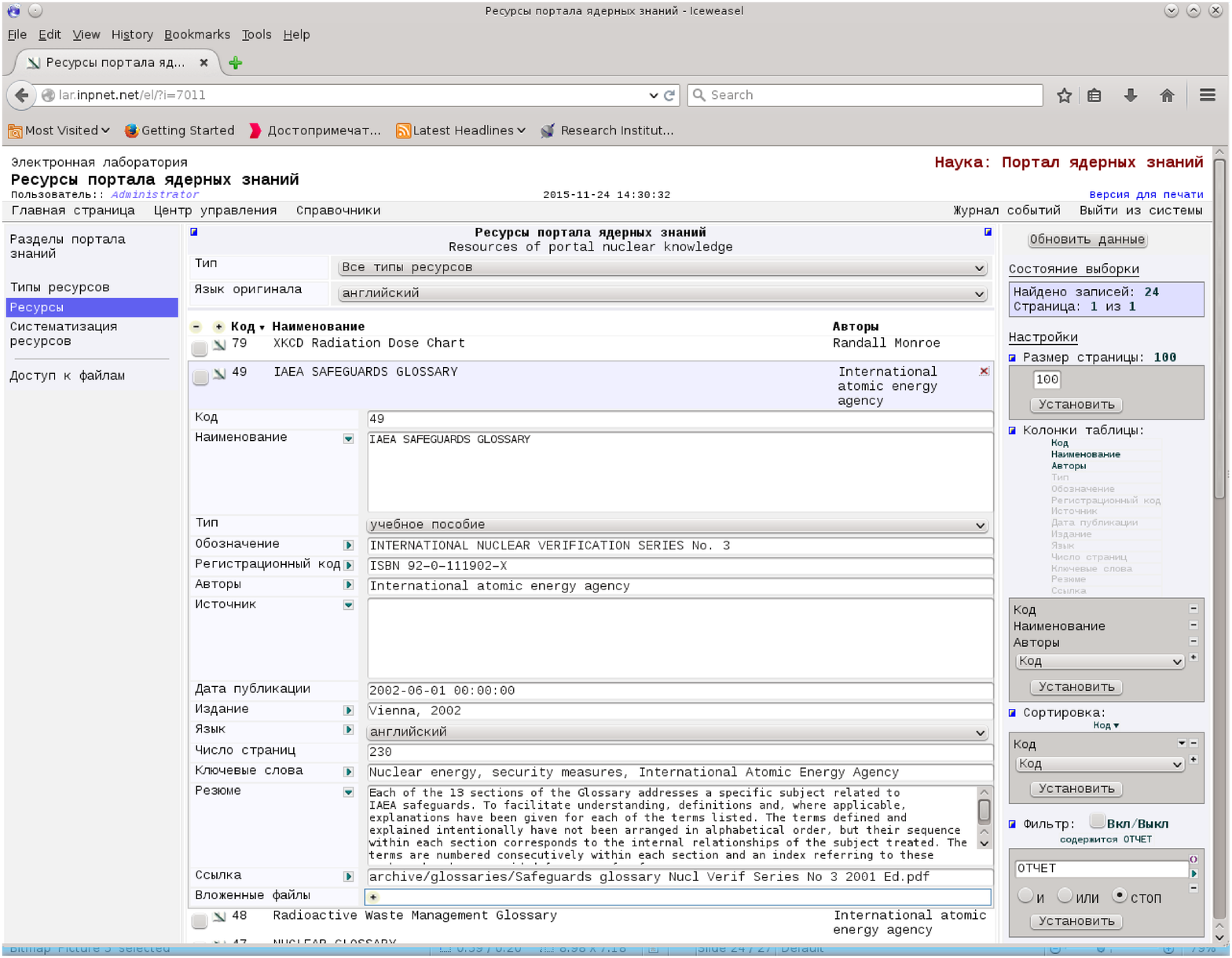}}
 \caption{Editing of BelNET resources}\label{fig5}}
\end{figure}

\begin{figure}[htb]\centering{
{\includegraphics[scale=0.27]{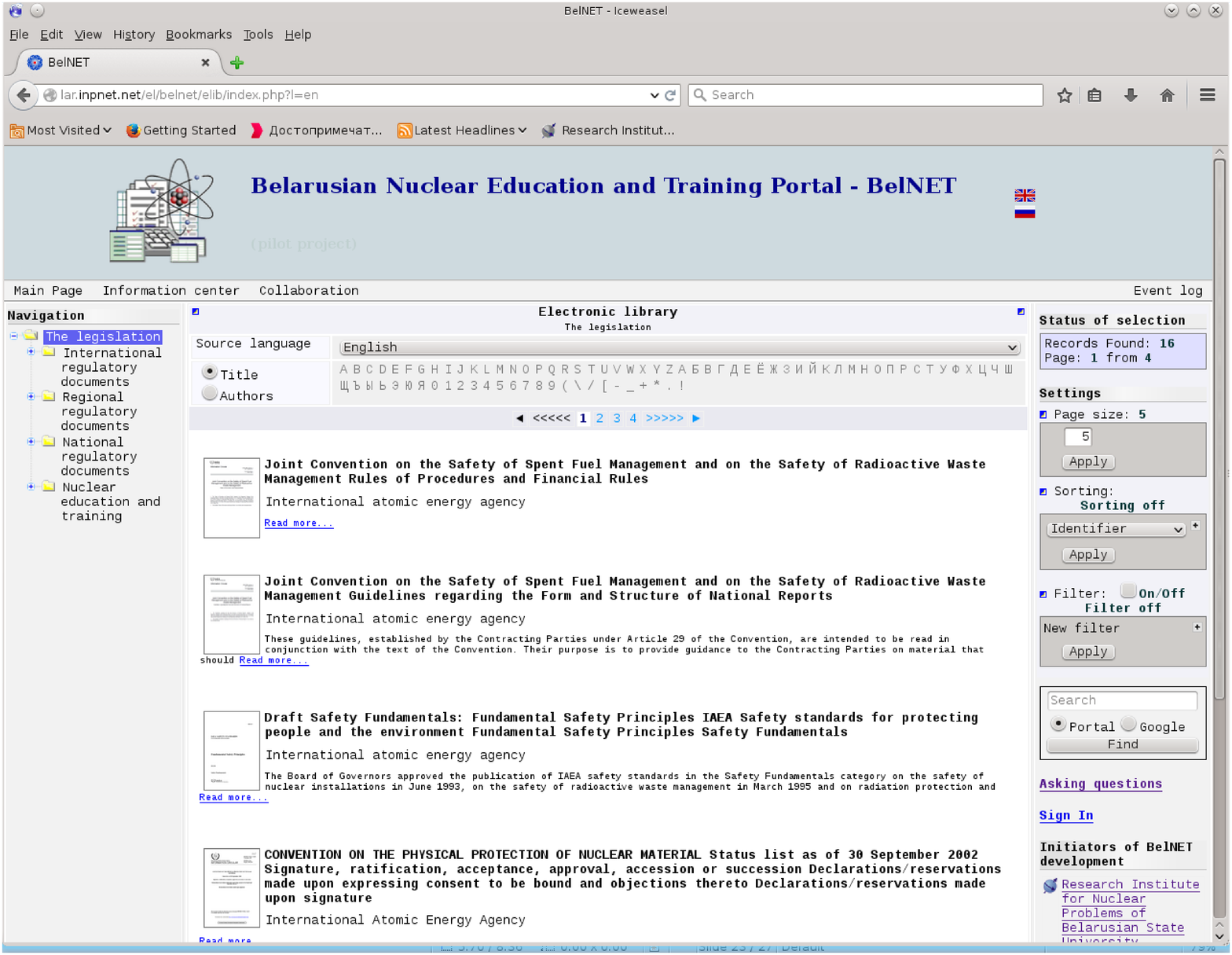}}
 \caption{BelNET- Informations center - Legislation}\label{fig6}}
\end{figure}

At present, BelNET software implements all basic functions of the
portal, including the ability to remotely (via the Internet) open
content editing, different sorting functions, filters, etc. (see
Fig.\ref{fig5}--\ref{fig6}). So, we can say that in the frame of
this work the original content management system (CMS) was
created. It includes the possibility to input texts and formulae
in LaTeX-similar form and load different types of files,
references, video, photos and pictures. On the basis of this CMS
educational and scientific portals of various profiles can be
created. Filling BelNET with information is underway.

\section {Lab practice for students}
We emphasize that filling the portal with information as well as
developing special materials for a distance learning system is a
time-consuming and long process. In this sense, the work on BelNET
is at the beginning.

Currently, a glossary of nuclear physics runnibg term and lab
practice for students are developed. Definitions included in the
glossary are formed on the basis of traditions prevailing in the
scientific community, national and interstate regulatory documents
with the account of the IAEA recommendations and the views of
draftsmen. The articles contain term definitions, area of
application and, if appropriate, analysis of differences between
the given definitions and common or standard ones.

In high school, nuclear physics is traditionally presented only by
a small theoretical section, which does not provide for
implementation of lab work because the sources of ionizing
radiation are forbodden by sanitary norms. However, the practical
skills that students receive in performing lab work allow them to
better understand the characteristics of ionizing radiation
passing through matter and the dangers associated with the use of
radioactive substances and principles of radiation protection.
This is very important because of ionizing radiation and
radioactive sources are widely used in medicine, engineering and
other areas. Now requirements for basic knowledge of radiation
principles and its impact on the environment are high in the light
of development of nuclear industry in Belarus. The necessity to
the general public of at least a minimum level of knowledge in
this area was confirmed by the history of the Chernobyl disaster.

As a part of BelNET content, it was decided to develop a series of
on-line guides for lab work on ionizing radiation passing through
matter. These lab works are oriented to high school and university
students as well as to anyone interested in this topic. The
general part of the series is an introduction with brief
information on nuclear physics and nuclear spectrometry. It
includes the desctiption of main features of the phenomenon of
radioactivity. The work of detectors of ionizing radiation is
explained as well as the principles of formation and
interpretation of experimental energy radiation spectra. The
formulae for estimating statistical errors of the experiment are
given. Practical part includes five lab works: "Determination of
the activity of radioactive source by a relative method",
"Absorption of electrons in matter", "Absorption of gamma rays in
matter", "Study of the penetrating power of gamma rays of
different energies", "Natural decay chains". Each practice
includes a brief description of the studied processes, necessary
for understanding the experimental part of the work, as well as
analysis of the obtained results. As a separate section of each
work, a guide on the order of experimental data processing,
calculation and analysis of finite quantities is given. The
experimental data (energy spectra of specific ionizing radiation)
obtained using the spectrometer of ionizing radiation of the
Department of Nuclear Physics of Belarusian State University are
available in the form of text files. Video files demonstrate the
spectrum set-up. This allows performing the lab work using only a
computer with a standard set of programs. Using a computer
calculator (eg. MS Excel), one can process the experimental
spectra, calculate the necessary values and present the obtained
results in graphical form. At the end of the series a test program
is given in order to check the correctness of obtained results, as
well as the level of understanding of the studied processes by the
user and his willingness to use the results, for example, to
estimate the parameters, necessary for protection against ionizing
radiation.

\section {Conclusion}

Belarusian educational and research portal of nuclear knowledge
BelNET is developed by the efforts of the best experts and
professorate in the field of nuclear knowledge of the Republic of
Belarus. The aim of the work is the promotion of nuclear
knowledge, the formation positive image of nuclear science and
attraction here the most able young people.
Created original CMS allows developing educational and scientific
portals of various profiles.

\end{document}